\documentclass{article}
\usepackage{dcase2022,amsmath,url,times,booktabs,tabularx,amsmath,amssymb}
\usepackage{array,multirow}
\usepackage{arydshln}
\usepackage{multicol}
\usepackage{footnote}
\usepackage{float}
\usepackage{cite}
\usepackage{enumitem}
\usepackage{color,soul}

\usepackage[pdftex]{graphicx}

\newcolumntype{C}[1]{>{\centering\arraybackslash}p{#1}}
\newcolumntype{L}[1]{>{\raggedright\arraybackslash}p{#1}}
\newcolumntype{R}[1]{>{\raggedleft\arraybackslash}p{#1}}

\newcommand{\vect}[1]{{\mbox{\boldmath $#1$}}}

\newcommand{\ReduceSpaceUnderFigure}{\vspace{0pt}}
\newcommand{\ReduceSpaceUnderTable}{\vspace{-0pt}}

\title{DESCRIPTION AND DISCUSSION ON DCASE 2022 CHALLENGE TASK 2: UNSUPERVISED ANOMALOUS SOUND DETECTION FOR MACHINE CONDITION MONITORING APPLYING DOMAIN GENERALIZATION TECHNIQUES}

\name{
Kota Dohi$^{1}$,
Keisuke Imoto$^{2}$,
Noboru Harada$^{3}$,
Daisuke Niizumi$^{3}$,
Yuma Koizumi$^{4}$,
Tomoya Nishida$^{1}$,
}
\secondlinename{
Harsh Purohit$^{1}$,
Takashi Endo$^{1}$,
Masaaki Yamamoto$^{1}$,
and Yohei Kawaguchi$^{1}$
}
\address{
$^1$ Hitachi, Ltd., Japan, \url{kota.dohi.gr@hitachi.com}\\
$^2$ Doshisha University, Japan, \url{keisuke.imoto@ieee.org}\\
$^3$ NTT Corporation, Japan, \url{noboru.harada.pv@hco.ntt.co.jp}\\
$^4$ Google, Japan, \url{koizumiyuma@google.com}\\
}

\begin{document}

\ninept
\maketitle

\begin{sloppy}

  \begin{abstract}

    We present the task description and discussion on the results of the DCASE 2022 Challenge Task 2: ``Unsupervised anomalous sound detection (ASD) for machine condition monitoring applying domain generalization techniques''.
    Domain shifts are a critical problem for the application of ASD systems. Because domain shifts can change the acoustic characteristics of data, a model trained in a source domain performs poorly for a target domain.
    In DCASE 2021 Challenge Task 2, we organized an ASD task for handling domain shifts. In this task, it was assumed that the occurrences of domain shifts are known. However, in practice, the domain of each sample may not be given, and the domain shifts can occur implicitly.
    In 2022 Task 2, we focus on domain generalization techniques that detects anomalies regardless of the domain shifts. Specifically, the domain of each sample is not given in the test data and only one threshold is allowed for all domains.
    Analysis of 81 submissions from 31 teams revealed two remarkable types of domain generalization techniques: 1) domain-mixing-based approach that obtains generalized representations and 2) domain-classification-based approach that explicitly or implicitly classifies different domains to improve detection performance for each domain.

  \end{abstract}


  \section{Introduction}
  \label{sec:intro}

  Anomalous sound detection (ASD)~\cite{koizumi2017neyman, kawaguchi2017how, koizumi2019neyman, kawaguchi2019anomaly, koizumi2019batch, suefusa2020anomalous, purohit2020deep} is the task of identifying whether the sound emitted from a target machine is normal or anomalous. Automatic detection of mechanical failure is essential in the fourth industrial revolution, which involves artificial intelligence (AI)--based factory automation. Prompt detection of machine anomalies by observing  sounds is useful for machine condition monitoring.

  One challenge regarding the application scope of ASD systems is that anomalous samples for training can be insufficient both in number and type. In 2020, we organized the fundamental ASD task in Detection and Classification of Acoustic Scenes and Event (DCASE) Challenge 2020 Task 2~\cite{Koizumi2020dcase}; ``\textit{unsupervised ASD}'' that was aimed to detect unknown anomalous sounds using only normal sound samples as the training data~\cite{koizumi2017neyman, kawaguchi2017how, koizumi2019neyman, kawaguchi2019anomaly, koizumi2019batch, suefusa2020anomalous, purohit2020deep}. For the wide spread application of ASD systems, advanced tasks such as handling of domain shifts should be tackled.
  Domain shifts are differences in acoustic characteristics between the source and target domain data caused by differences in a machine's operational conditions or environmental noise. Because these shifts are caused by factors other than anomalies, the detection performance of models trained with the source domain data can degrade for the target domain data.
  Therefore, in 2021, we organized DCASE Challenge 2021 Task 2~\cite{Kawaguchi2021}, ``\textit{unsupervised ASD under domain shifted conditions}'' that focused on handling domain shifts using domain adaptation techniques.

  The task in 2021 involved the use of domain adaptation techniques under two assumptions. First, all domain shifts have been detected in advance, and the domain of each sample is known. Second, the domain shifts do not occur too frequently for the model to adapt. However, these assumptions may not hold for certain real-world scenarios. For example, a machine's background sound can be affected by various sound sources surrounding the machine, and it can be difficult to identify the cause of changes and attribute the changes to the domain shift. Also, because the operational conditions of the machine can change within a short period, adapting the model every time can be too costly.
  Therefore, methods have to be investigated such that the detection of domain shifts is unnecessary and frequent occurrences of domain shifts can be handled.

  To solve the problem described above, we designed DCASE challenge 2022 Task 2 ``\textit{Unsupervised Detection of Anomalous Sounds for Machine Condition Monitoring Applying Domain Generalization Techniques}''. This task is aimed at developing domain generalization techniques to handle domain shifts. The task involves the use of domain generalization techniques so that the developed ASD systems do not require detection of the domain shifts or adaptation of the model. Specifically, to evaluate the generalization performance, the domain of each sample is not provided in the test data. To enhance generalization of the model, attributes that caused domain shifts are also provided in the training data.

  We received 81 submissions from 31 teams. By analyzing these submissions, we found two types of domain generalization techniques: 1) domain-mixing-based approach and 2) domain-classification-based approach. The domain-mixing-based approach aims at obtaining generalized representations across domains by mixing data from different domains. In contrast, the domain-classification-based approach differentiates different domains so that the model can be specialized for each domain.

  \section{Unsupervised Anomalous Sound Detection Applying Domain Generalization Techniques}
  \label{sec:uasd}

  Let the $L$-sample time-domain observation $\vect{x} \in \mathbb{R}^L$ be an audio clip that includes a sound emitted from a machine. The ASD task is a task to determine whether a machine is in a normal or anomalous state using an anomaly score $\mathcal{A}_{\theta}(\vect{x})$ calculated by an anomaly score calculator $\mathcal{A}: \mathbb{R}^L \to \mathbb{R}$ with parameters $\theta$.
  %The anomaly score is large for anomalous samples and small for normal samples. 
  %The input of $\mathcal{A}$ is the audio clip $\vect{x}$ or $\vect{x}$ with additional information, and the output is an  $\mathcal{A}_{\theta}(\vect{x})$. 
  The machine is determined to be anomalous when  $\mathcal{A}_{\theta}(\vect{x})$ exceeds a pre-defined threshold $\phi$ as
  \begin{equation}
    \mbox{Decision} = \left\{
    \begin{array}{ll}
      \mbox{Anomaly} & (\mathcal{A}_{\theta}(\vect{x}) > \phi) \\
      \mbox{Normal}  & (\mbox{otherwise}).
    \end{array}
    \right.
    \label{eq:det}
  \end{equation}
  The primary difficulty in this task is to train $\mathcal{A}$
  %so that $\mathcal{A}_{\theta}(\vect{x})$ becomes large when the machine is anomalous, 
  with only normal sounds. This is because anomalies are rarely obtained in practice.

  Domain-shift is another major issue in real-world applications.
  % In addition to unsupervised ASD, we have to solve the domain-shift problem in real-world cases.
  Domain shifts mean a difference in conditions between training and testing. The conditions are machine's operational conditions such as its speed, load, and temperature, or the environmental conditions such as the type of environmental noise, level of the noise, and location of the microphone. Differences in these conditions change the distribution of data and degrades the detection performance.
  Let us define two domains: \textbf{source domain} and \textbf{target domain}, where the source domain is the original condition with enough training clips and the target domain is another condition with zero or a few training clips. Also, let $\mathcal{D}_S$, $\mathcal{D}_T$, $\mathcal{D}_{SA}$, and $\mathcal{D}_{TA}$ be the distributions of $\vect{x}$ under the normal condition in the source domain, normal condition in the target domain, anomalous condition in the source domain, and anomalous condition in the target domain, respectively.

  % The task in DCASE 2021 Task 2 was to either determine if $\vect{x_s}$ from the source domain is generated from $\mathcal{D}_S$ or $\mathcal{D}_{SA}$ using an anomaly score calculator $\mathcal{A}_{\theta_s}(\vect{x})$ and a threshold $\phi_{s}$, and determine if $\vect{x_t}$ from the target domain is generated from $\mathcal{D}_T$ or $\mathcal{D}_{TA}$ using an anomaly score calculator $\mathcal{A}_{\theta_t}(\vect{x})$ and a threshold $\phi_{t}$.
  The task in DCASE 2021 Task 2 involved two tasks. One was to detect anomalies in the source domain: determine whether $\vect{x_s}$ is from $\mathcal{D}_S$ or $\mathcal{D}_{SA}$ using an anomaly score calculator $\mathcal{A}_{\theta_s}(\vect{x})$ and a threshold $\phi_{s}$. The other was detection in the target domain: whether $\vect{x_t}$ is generated from $\mathcal{D}_T$ or $\mathcal{D}_{TA}$ using an anomaly score calculator $\mathcal{A}_{\theta_t}(\vect{x})$ and a threshold $\phi_{t}$.
  The task was set to develop domain adaptation techniques so that the detection performance on the target domain can be improved by adaptation on the model trained with the source domain data. Although this problem setting assumes that the domain (source/target) of each sample is known, in practice, the detection of domain shifts can be difficult and the domain may not be available. Also, the use of domain adaptation techniques can be too costly if the domain shifts occur too frequently.

  % We identified four types of real-world scenarios that illustrate these problems.\\
  We show four types of real-world scenarios for these problems.
  \vspace{-8pt}

  \noindent
  \textbf{Domain shifts due to differences in machine’s conditions} \\
  Characteristics of a machine sound can change due to changes in the machine’s operational conditions. Although these shifts can be detected, if these conditions change within a short period of time, it can be too costly to adapt the model every time.\\
  \vspace{-8pt}

  \noindent
  \textbf{Domain shifts due to differences in environmental conditions} \\
  Because characteristics of background noise can be affected by various factors, it is difficult to detect these shifts. Therefore, a model that is unaffected by these shifts is desirable.\\
  \vspace{-8pt}

  \noindent
  \textbf{Domain shifts due to maintenance} \\
  Characteristics of a machine sound can change after maintenance or parts replacement. Though these shifts can be detected, adapting the model every time can be costly.\\
  \vspace{-8pt}

  \noindent
  \textbf{Domain shifts due to differences in recording devices} \\
  In real-world scenarios, many microphones are installed at different locations, and these microphones may be from different manufacturers. Although these shifts can be detected, adapting the model for each location or microphone can be too costly.\\
  \vspace{-8pt}

  As a possible solution to handle these problems, domain generalization techniques should be investigated. Domain generalization techniques for ASD aims at detecting anomalies from different domains with a single threshold. These techniques, unlike domain adaptation techniques, do not require detection of domain shifts or adaptation of the model in the testing phase. Therefore, domain generalization techniques can be used for handling domain shifts that are difficult to detect or too costly to adapt.

  The DCASE 2022 Task 2 is set to develop domain generalization techniques for ASD.
  Because the domain generalization techniques are expected to work regardless of the domains, the domain of each sample is not given in the test data.
  The task is to determine if $\vect{x}$ is from the normal condition $\mathcal{D}_S \cup \mathcal{D}_T$ or anomalous condition $\mathcal{D}_{SA} \cup \mathcal{D}_{TA}$ using an anomaly score calculator $\mathcal{A}_{\theta}(\vect{x})$ and $\phi$. Because the differences in operational or environmental conditions make $\mathcal{D}_S \ne \mathcal{D}_T$, the decision must be executed without being affected by the differences between different domains.

  \section{Task Setup}
  \label{sec:task}

  \subsection{Dataset}
  \label{sec:dataset}

  We used ToyADMOS2~\cite{harada2021toyadmos2} and MIMII DG~\cite{Dohi2022} to generate the dataset. The dataset consists of normal/anomalous operating sounds from seven types of toy/real machines (ToyCar, ToyTrain, fan, gearbox, bearing, slide rail, and valve).

  Each recording is a single-channel and 10-sec-long audio with a sampling rate of 16 kHz. We mixed machine sounds recorded at laboratories and the environmental noise recorded at real-world factories to create the training/test data.
  Details of the recording procedure can be found in \cite{harada2021toyadmos2} and \cite{Dohi2022}.

  % In this task, we define two important terms: \textbf{machine type} and \textbf{section}.
  %\begin{itemize}
  %\item \textbf{Machine type} means the type of machine, which can be one of seven in this task: fan, gearbox, bearing, slide rail, ToyCar, ToyTrain, and valve.
  %\item \textbf{Section} is defined as a subset of the data within one machine type and corresponds to a type of domain shift scenario. 
  %It is also a unit for calculating performance metrics.
  %\end{itemize}
  In this dataset, \textbf{Machine type} means the type of machine.
  \textbf{Section} is defined as a subset of the data within a machine type and corresponds to a type of domain shift scenario.

  We provide three datasets:
  \textbf{development dataset}, \textbf{additional training dataset}, and \textbf{evaluation dataset}.
  The \textbf{development dataset} consists of three sections (Sections 00, 01, and 02), which are sets of the training and test data.
  Each section provides
  (i) 990 normal clips from a source domain for training,
  (ii) 10 normal clips from a target domain for training,
  (iii) 100 normal clips and 100 anomalous clips from both domains for the test.
  We provided domain information (source/target) in the test data for the convenience of participants.
  Attributes represent the operational or environmental conditions, e.g. velocity of slide rail and level of noise (SNR) mixed in fan data.
  %\item 
  The \textbf{additional training dataset} provides training clips for three sections (Sections 03, 04, and 05).
  Each section consists of
  (i) 990 normal clips in a source domain for training and
  (ii) 10 normal clips in a target domain for training.
  Attributes are also provided.
  The \textbf{evaluation dataset} provides test clips for three sections (Sections 03, 04, and 05).
  Each section consists of 200 test clips, none of which have a condition label (i.e., normal or anomaly) or the domain information. Attributes are not provided.
  %\end{itemize}
  The main difference from our task in 2021 is that the domain information is not given in the evaluation dataset.
  Thus, the participants have to develop a system that performs well regardless of the domains.

  \subsection{Evaluation metrics}
  \label{sec:metrics}

  This task is evaluated with the area under the receiver operating characteristic (ROC) curve (AUC) and the partial AUC (pAUC).
  The pAUC is calculated as the AUC over a low false-positive-rate (FPR) range $\left[ 0, p \right]$.
  % The reason for the additional use of the pAUC is based on practical requirements. If an ASD system frequently gives false alarms frequently, we cannot trust it. Therefore, it is important to increase the true-positive rate under low FPR conditions. 
  In this task, we used $p=0.1$.

  Because the domain generalization task requires detecting anomalies using the same threshold between domains, the pAUC has to be calculated for each section, not for each domain. We calculated the AUC for each domain and pAUC for each section as
  \begin{align}
    {\rm AUC}_{m, n, d} & = \frac{1}{N^{-}_{d}N^{+}_{n}} \sum_{i=1}^{N^{-}_{d}} \sum_{j=1}^{N^{+}_{n}}
    \mathcal{H} (\mathcal{A}_{\theta} (x_{j}^{+}) - \mathcal{A}_{\theta} (x_{i}^{-})),                  \\
    %&{\rm pAUC}_{m, n} = \frac{1}{\lfloor p N^{-}_{d} \rfloor N^{+}_{n}} \sum_{i=1}^{\lfloor p N^{-}_{d} \rfloor N^{+}_{n}} \sum_{j=1}^{N^{+}_{n}}
    {\rm pAUC}_{m, n}   & = \frac{1}{P^{-}_{n} N^{+}_{n}} \sum_{i=1}^{P^{-}_{n}} \sum_{j=1}^{N^{+}_{n}}
    \mathcal{H} (\mathcal{A}_{\theta} (x_{j}^{+}) - \mathcal{A}_{\theta} (x_{i}^{-})),
  \end{align}
  where
  $P^{-}_n = \lfloor p N^{-}_{n} \rfloor$, $m$ represents the index of a machine type,
  $n$ represents the index of a section,
  $d = \{ {\rm source}, {\rm target} \}$ represents a domain,
  $\lfloor \cdot \rfloor$ is the flooring function,
  and $\mathcal{H} (x)$ returns 1 when $x > 0$ and 0 otherwise.
  Here, $\{\mathcal{A}_{\theta} (x^{-}_{i})\}$ and $\{\mathcal{A}_{\theta} (x^{+}_{j})\}$ are sets of anomaly scores of normal and anomalous test clips, ordered in descending power, respectively.
  $N^{-}_{d}$ is the number of normal test clips in domain $d$,
  $N^{-}_{n}$ and $N^{+}_{n}$ are the number of normal and anomalous test clips in section $n$, respectively. We calculated ${\rm AUC}_{m, n, d}$ to evaluate the contribution of each domain to ${\rm AUC}_{m, n}$, as it holds that ${\rm AUC}_{m, n}= \sum_{d} {\rm AUC}_{m, n, d}$
  if $N^{-}_{source} = N^{-}_{target}$.

  The official score $\Omega$ for ranking submitted systems is given by the harmonic mean of the AUC and pAUC scores over all machine types and sections as follows:
  \begin{eqnarray}
    \Omega &=& h \left\{ {\rm AUC}_{m, n, d}, \ {\rm pAUC}_{m, n} \quad | \quad \right. \nonumber \\
    && \left. m \in \mathcal{M}, \  n \in \mathcal{S}(m), \ d \in \{ {\rm source}, {\rm target} \} \right\},
  \end{eqnarray}
  where $h\left\{\cdot\right\}$ represents the harmonic mean (over all machine types, sections, and domains), $\mathcal{M}$ represents the set of machine types, and $\mathcal{S}(m)$ represents the set of sections for machine type $m$.

  % As the equations above show, a threshold does not need to be determined to calculate AUC, pAUC, or the official score because it is the anomaly scores of normal test clips. In addition, because the threshold must be determined in real-world applications,
  % participants are also required to submit the normal/anomaly decision results.
  Participants are required to submit the anomaly score and normal/anomaly decision result of each test clip. Even though the official score can be calculated with only the anomaly scores, decision results are also required because we must determine the threshold in real-world applications.
  % In contrast to previous years' task, participants are also required to submit the normal/anomaly decision results.
  % This is because of that, in real applications, the threshold must be determined, and a decision must be made as to whether it is normal or abnormal. 

  \subsection{Baseline systems and results}
  \label{sec:baseline}

  The task organizers provide
  an autoencoder (AE)-based
  and
  a MobileNetV2-based baseline systems.
  % Since these models are the same as previous years baseline systems~\cite{Koizumi2020dcase,Kawaguchi2021},
  % we briefly introduce these systems due to limitations of the space.
  %We present these baseline systems and their results.
  %\subsubsection{Autoencoder-based baseline}
  %\label{sec:ae}

  % The first baseline system is based on the autoencoder (AE), which is the same as in DCASE 2020 Challenge Task2~\cite{Koizumi2020dcase} and DCASE 2021 Challenge Task2~\cite{Kawaguchi2021}.
  The AE-based system calculates the anomaly score as the reconstruction error of the sound.
  To determine the threshold, we assume that anomaly scores %$\mathcal{A}_{\theta} (\vect{x})$
  of normal sound follows a gamma distribution.
  %$\mathcal{A}_{\theta}$
  % anomaly scores of [WHAT?] follows a gamma distribution. 
  % The parameters of the gamma distribution are estimated from anomaly scores of training samples.
  % anomaly scores of normal training samples,
  The parameters of the gamma distribution are estimated from the anomaly scores of normal sound in the training data,
  and the threshold is calculated by the 90th percentile of the gamma distribution.
  A test clip is determined to be anomalous if its anomaly score exceeds the threshold.
  In the MobileNetV2-based system \cite{giri2020self, primus2020anomalous, inoue2020detection}, classifiers such as the MobileNetV2~\cite{sandler2018mobilenetv2} are trained to identify from which section the observed signal was generated.
  %In other words, it outputs the softmax value that is the predicted probability for each section. 
  The anomaly score is calculated as the averaged negative logit of the predicted probabilities for the correct section.
  %
  %The acoustic feature (two-dimensional image) $\psi_t \in \mathbb{R}^{P \times F}$ is calculated in the same manner as for the AE-based baseline. By shifting the context window by $L$ frames, $B (= \lfloor \frac{T - P}{L} \rfloor)$ images are extracted. The frame size of STFT is 64ms, and the hop size is 50 \%. Also, $F=128$, $P=64$, and $L=8$.
  %The ADAM optimizer is used, and we fix the learning rate to 0.00001. We stop the training process after 20 epochs, and the batch size is 32. We train models independently for each machine type using normal clips from all sections of that machine type. The anomaly score is calculated as
  %\begin{equation}
  %\mathcal{A}_{\theta}(X) = \frac{1}{B} \sum_{b = 1}^B \log \left( \frac{1 - p_{\theta}(\psi_{t(b)})}{p_{\theta}(\psi_{t(b)})} \right),
  %\end{equation}
  %where $t(b)$ is the beginning frame index of the $b$-th image, and $p_{\theta}$ is the softmax output by MobileNetV2 for the correct section.
  The threshold is calculated in the same manner as in the AE-based baseline.

  Tables \ref{tab:ae_results} and \ref{tab:mob_results} show the AUC and pAUC  for the two baselines.
  Because the results produced with a GPU are generally non-deterministic, the average and standard deviations from five independent trials are also shown in the tables.
  \vspace{-8pt}

  %\subsubsection{Results}
  %\label{sec:results}

  \setlength{\tabcolsep}{1mm}
  \begin{table}[t]
    \begin{center}
      \caption{Results of the AE-based baseline}
      \label{tab:ae_results}
      \scriptsize
      \begin{tabular}{l l l l p{1pt} l }
        \hline
        \multicolumn{2}{@{}l}{\multirow{2}{*}{ \begin{tabular}{@{\hskip0pt}l@{\hskip0pt}} Section \end{tabular} }} &
        \multicolumn{2}{c}{AUC [\%]}                                        &    &
        \multicolumn{1}{c}{pAUC [\%]}                                                                                                           \\
        \cline{3-4}
                                                                            &    &
        \multicolumn{1}{c}{Source}                                          &
        \multicolumn{1}{c}{Target}                                          &    &                                                              \\
        \hline
        \multirow{3}{*}{ToyCar}
                                                                            & 00 & $86.42 \pm 1.10$  & $41.48 \pm 6.11$  &  & $51.31 \pm 1.34$  \\
                                                                            & 01 & $89.85 \pm 1.39$  & $41.93 \pm 5.36$  &  & $54.08 \pm 1.84$  \\
                                                                            & 02 & $98.84 \pm 0.52$  & $26.50 \pm 13.52$ &  & $52.79 \pm 1.04$  \\
        \hline

        \multirow{3}{*}{ToyTrain}
                                                                            & 00 & $67.54 \pm 0.97$  & $33.68 \pm 3.12$  &  & $52.72 \pm 1.63$  \\
                                                                            & 01 & $79.32 \pm 0.82$  & $29.87 \pm 5.62$  &  & $50.64 \pm 2.33$  \\
                                                                            & 02 & $84.08 \pm 0.38$  & $15.52 \pm 14.90$ &  & $48.33 \pm 2.33$  \\
        \hline

        \multirow{3}{*}{Bearing}
                                                                            & 00 & $67.85 \pm 19.61$ & $60.17 \pm 7.24$  &  & $54.41 \pm 5.72$  \\
                                                                            & 01 & $59.67 \pm 12.67$ & $64.65 \pm 12.63$ &  & $55.09 \pm 3.36$  \\
                                                                            & 02 & $61.71 \pm 33.52$ & $60.55 \pm 35.10$ &  & $64.18 \pm 19.79$ \\
        \hline

        \multirow{3}{*}{Fan}
                                                                            & 00 & $84.69 \pm 1.74$  & $39.35 \pm 9.35$  &  & $59.95 \pm 2.00$  \\
                                                                            & 01 & $71.69 \pm 0.69$  & $44.74 \pm 1.79$  &  & $51.12 \pm 0.55$  \\
                                                                            & 02 & $80.54 \pm 1.42$  & $63.49 \pm 2.36$  &  & $62.88 \pm 1.55$  \\
        \hline

        \multirow{3}{*}{Gearbox}
                                                                            & 00 & $64.63 \pm 0.88$  & $64.79 \pm 1.06$  &  & $60.93 \pm 2.31$  \\
                                                                            & 01 & $67.66 \pm 0.51$  & $58.12 \pm 0.38$  &  & $53.74 \pm 0.56$  \\
                                                                            & 02 & $75.38 \pm 0.75$  & $65.57 \pm 0.82$  &  & $61.51 \pm 0.69$  \\
        \hline

        \multirow{3}{*}{Slide rail}
                                                                            & 00 & $81.92 \pm 0.81$  & $58.04 \pm 1.22$  &  & $61.65 \pm 1.22$  \\
                                                                            & 01 & $67.85 \pm 0.53$  & $50.30 \pm 1.25$  &  & $53.06 \pm 0.53$  \\
                                                                            & 02 & $86.66 \pm 0.39$  & $38.78 \pm 5.13$  &  & $53.44 \pm 1.18$  \\
        \hline

        \multirow{3}{*}{Valve}
                                                                            & 00 & $54.24 \pm 0.68$  & $52.73 \pm 1.93$  &  & $52.15 \pm 0.25$  \\
                                                                            & 01 & $50.45 \pm 3.67$  & $53.01 \pm 1.73$  &  & $49.78 \pm 0.19$  \\
                                                                            & 02 & $51.56 \pm 2.89$  & $43.84 \pm 1.11$  &  & $49.24 \pm 0.65$  \\
        \hline
      \end{tabular}
    \end{center}
    \ReduceSpaceUnderTable
    % \vspace{-1pt}
  \end{table}

  \setlength{\tabcolsep}{1mm}
  \begin{table}[htpb]
    \begin{center}
      \caption{Results of the MobileNetV2-based baseline}
      \label{tab:mob_results}
      \scriptsize
      \begin{tabular}{l l l l p{1pt} l }
        \hline
        \multicolumn{2}{@{}l}{\multirow{2}{*}{ \begin{tabular}{@{\hskip0pt}l@{\hskip0pt}} Section \end{tabular} }} &
        \multicolumn{2}{c}{AUC [\%]}                                        &    &
        \multicolumn{1}{c}{pAUC [\%]}                                                                                                           \\
        \cline{3-4}
                                                                            &    &
        \multicolumn{1}{c}{Source}                                          &
        \multicolumn{1}{c}{Target}                                          &    &                                                              \\
        \hline
        \multirow{3}{*}{ToyCar}
                                                                            & 00 & $47.40 \pm 7.22$  & $56.40 \pm 4.11$  &  & $49.96 \pm 2.56$  \\
                                                                            & 01 & $62.02 \pm 11.07$ & $56.38 \pm 11.31$ &  & $50.92 \pm 2.52$  \\
                                                                            & 02 & $74.19 \pm 7.94$  & $45.64 \pm 11.32$ &  & $56.51 \pm 6.07$  \\
        \hline

        \multirow{3}{*}{ToyTrain}
                                                                            & 00 & $46.02 \pm 12.21$ & $49.41 \pm 15.14$ &  & $50.25 \pm 1.49$  \\
                                                                            & 01 & $71.96 \pm 5.72$  & $45.14 \pm 13.66$ &  & $52.97 \pm 4.61$  \\
                                                                            & 02 & $63.23 \pm 25.60$ & $44.34 \pm 21.50$ &  & $51.54 \pm 4.34$  \\
        \hline

        \multirow{3}{*}{Bearing}
                                                                            & 00 & $67.85 \pm 19.61$ & $60.17 \pm 7.24$  &  & $54.41 \pm 5.72$  \\
                                                                            & 01 & $59.67 \pm 12.67$ & $64.65 \pm 12.63$ &  & $55.09 \pm 3.36$  \\
                                                                            & 02 & $61.71 \pm 33.52$ & $60.55 \pm 35.10$ &  & $64.18 \pm 19.79$ \\
        \hline

        \multirow{3}{*}{Fan}
                                                                            & 00 & $71.07 \pm 19.84$ & $62.13 \pm 12.50$ &  & $55.40 \pm 11.29$ \\
                                                                            & 01 & $76.26 \pm 4.95$  & $35.12 \pm 13.38$ &  & $52.14 \pm 4.08$  \\
                                                                            & 02 & $67.29 \pm 10.34$ & $58.02 \pm 7.46$  &  & $65.14 \pm 1.09$  \\
        \hline

        \multirow{3}{*}{Gearbox}
                                                                            & 00 & $63.54 \pm 9.46$  & $67.02 \pm 13.50$ &  & $62.12 \pm 11.66$ \\
                                                                            & 01 & $66.68 \pm 12.29$ & $66.96 \pm 8.92$  &  & $56.85 \pm 4.47$  \\
                                                                            & 02 & $80.87 \pm 7.85$  & $43.15 \pm 16.12$ &  & $50.62 \pm 7.73$  \\
        \hline

        \multirow{3}{*}{Slide rail}
                                                                            & 00 & $87.15 \pm 2.71$  & $80.77 \pm 4.53$  &  & $71.57 \pm 5.28$  \\
                                                                            & 01 & $49.66 \pm 30.46$ & $32.07 \pm 46.84$ &  & $48.21 \pm 2.73$  \\
                                                                            & 02 & $72.70 \pm 11.67$ & $32.94 \pm 19.77$ &  & $49.69 \pm 1.63$  \\
        \hline

        \multirow{3}{*}{Valve}
                                                                            & 00 & $75.26 \pm 4.84$  & $43.60 \pm 14.38$ &  & $55.37 \pm 5.86$  \\
                                                                            & 01 & $54.78 \pm 5.37$  & $60.43 \pm 5.08$  &  & $54.69 \pm 3.87$  \\
                                                                            & 02 & $76.26 \pm 1.02$  & $78.74 \pm 2.64$  &  & $85.74 \pm 0.08$  \\
        \hline
      \end{tabular}
    \end{center}
    \ReduceSpaceUnderTable
  \end{table}

  \section{Challenge Results}
  \label{sec:results}

  \subsection{Results for evaluation dataset}

  We received 81 submissions from 31 teams, and 22 teams outperformed the MobileNetV2-based baseline in the official score. In Figure \ref{fig:aucs}, the harmonic means of the AUCs are shown for top 10 teams \cite{LiuCQUPT2022, KuroyanagiNU-HDL2022, GuanHEU2022, DengTHU2022, VenkateshMERL2022, WeiHEU2022, MoritaSECOM2022, BaiJLESS2022, VerbitskiyDS2022, WilkinghoffFKIE2022}. Although the AUCs change drastically between different machine types and teams,
  these highly ranked teams outperformed the baselines for most of the machine types. It is worth noting that, for these teams, the source-domain AUC
  did not correlate with the official rank (correlation coefficient was $-0.033$) while the target-domain AUC did (correlation coefficient was $-0.862$). This indicates that handling domain shifts and generalizing the model was the key to better ranks among highly ranked teams.

  \begin{figure*}[t]
    \begin{center}
      \includegraphics[width=1.0\hsize,clip]{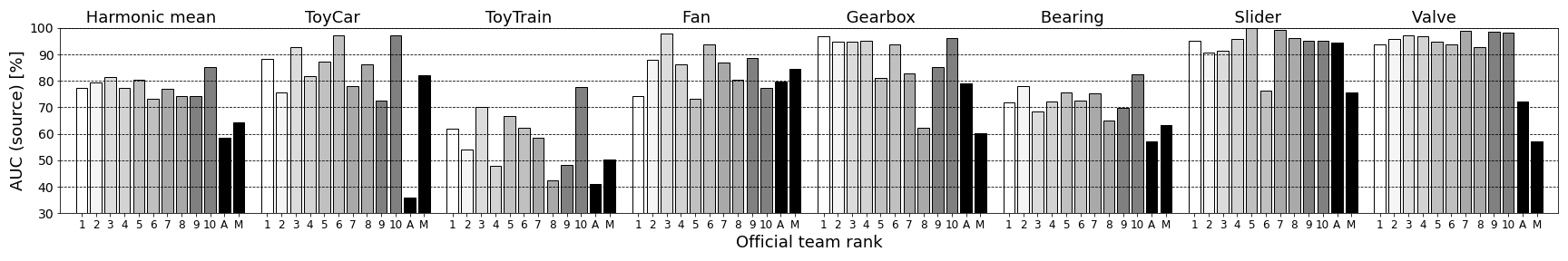}\\
      \includegraphics[width=1.0\hsize,clip]{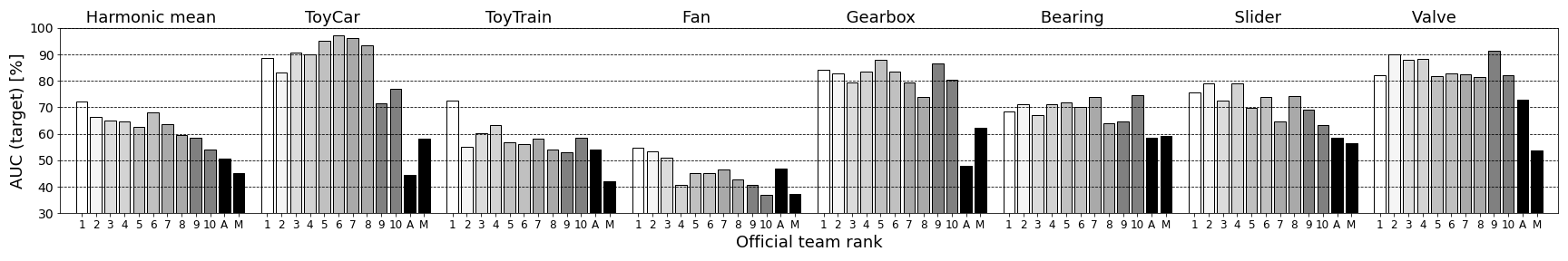}\\
      \caption{Evaluation results of top 10 teams in the ranking. Average source-domain AUC (Top) and target-domain AUC
        (bottom) for each machine type. Label ``A'' and ``M'' on the x-axis denote AE-based and MobileNetV2-based baselines,
        respectively.}
      \label{fig:aucs}
    \end{center}
    \ReduceSpaceUnderFigure
  \end{figure*}

  \begin{figure}[t]
    \begin{center}
      \includegraphics[width=0.9\hsize,clip]{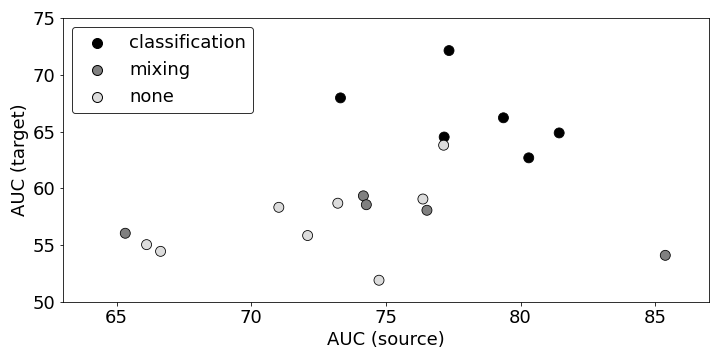}\\
      \vspace{-13pt}
      \caption{Average source-domain AUC and target-domain AUC of the top 20 teams. ``classification'' denotes teams that used domain-classification-based approaches, ``mix-up'' denotes teams that used domain-mixing-based approaches, and ``none'' denotes teams that did not use particular domain generalization techniques.}
      \label{fig:types}
    \end{center}
    \ReduceSpaceUnderFigure
  \end{figure}

  We find that domain generalization approaches adopted by the participants can be categorized into two types: domain-mixing-based approach and domain-classification-based approach. These methods achieved the aim of the task by generalizing the model using data with different attributes.
  Figure \ref{fig:types} shows the average source-domain and target-domain AUC of the top 20 teams. Domain-classification-based approaches outperformed other approaches especially for the target domain. However, these approaches may be specialized for the types of target domain data provided in both the training and test data, and thus may not perform well for those not included in the train data.
  We describe the details in the following.

  \subsection{Domain-mixing-based approach}
  Domain-mixing-based approach extracts common representations between domains. These include batch mixing that use data from both domains in a batch to train a model \cite{KuroyanagiNU-HDL2022, BaiJLESS2022}, Mixup \cite{zhang2018mixup} that synthesizes data from both domains to obtain intermediate representations \cite{KuroyanagiNU-HDL2022, BaiJLESS2022, WilkinghoffFKIE2022, NejjarETH2022}, and data augmentation techniques to obtain robust representations \cite{VerbitskiyDS2022}. These techniques use the target domain data to expand the normal conditions for the model so that the model can be generalized to better handle domain shifts. However, as shown in Figure \ref{fig:types}, they have been outperformed by domain-classification-based approaches. This can be that, for the Mixup and data augmentation, synthesized data was not useful for representing the target domain data. One future direction can be on obtaining meaningful synthetic representations with the aid of external information such as the attribute information.

  \subsection{Domain-classification-based approach}
  Domain-classification-based approach distinguishes the source and target domain data to obtain better detection performance for each domain. The 1st and 6th place teams \cite{LiuCQUPT2022, WeiHEU2022} used distances between the embedding of a domain and that of the test data to calculate anomaly scores. Because the domain of each sample can be estimated by the domain with shorter distance, this approach can be regarded as implicitly classifying the domain of each sample. The 2nd, 3rd, 4th, and 5th place teams \cite{KuroyanagiNU-HDL2022, GuanHEU2022, DengTHU2022, VenkateshMERL2022} explicitly trained a classifier to distinguish the attributes or the domains. The 5th place teams trained an attribute classifier and a section classifier so that both the domain-wise information from the attribute classifier and the domain-independent information from the section classifier can be obtained.

  As shown in Figure \ref{fig:types}, the domain-classification-based approach outperformed the domain-mixing-based approach.
  This can be because the normal conditions are defined for each specific domain, unlike the domain-mixing-based approach that defines normal conditions over all domains.
  However, this approach assumes that the target domain data in the train data includes all types of the target domain data in the test data. If the target domain data in the test data contains too many types of data not included in the train data, the classifier may fail to distinguish domains, which can degrade the detection performance. Therefore, further investigation is needed to examine the ability of this approach to handle completely unseen target domain data.

  \section{Conclusion}
  This paper presented an overview of the task and analysis of the solutions submitted to DCASE 2022 Challenge Task 2. To handle domain shifts that occur implicitly, the task was dedicated to developing domain generalization techniques. %The aim of the task was achieved by 81 submissions from 31 teams.
  %Analysis of 81 submissions from 31 teams 
  The organization of the task revealed two approaches that can be useful for domain generalization task: domain-mixing-based approach and domain-classification-based approach. For the former approach, obtaining more meaningful synthetic representations from multiple domains is left for future works. For the latter approach, future works can focus on analyzing the effect of this approach on completely unseen types of target domain data.

  \clearpage
  \setlength{\itemsep}{-0.0pt}
  \setlength{\baselineskip}{10.0pt}
  \bibliographystyle{IEEEtran}
  \footnotesize{
    \bibliography{refs}
  }

\end{sloppy}
\end{document}